\documentclass[twocolumn,aps,prb,superscriptaddress,amssymb,showpacs]{revtex4}
\usepackage[dvips,final]{graphicx}
\begin{document}
\bibliographystyle{apsrev}
\title{Mean field superconductivity approach in two dimensions: Hydrogen in Graphite} 
\author{N. Garc\'ia}\email{nicolas.garcia@fsp.csic.es}
\affiliation{Division of Superconductivity and Magnetism, Institut
f\"ur Experimentelle Physik II, Universit\"{a}t Leipzig,
Linn\'{e}stra{\ss}e 5, D-04103 Leipzig, Germany}
\affiliation{Laboratorio de F\'isica de Sistemas Peque\~nos y
Nanotecnolog\'ia,
 Consejo Superior de Investigaciones Cient\'ificas, E-28006 Madrid, Spain}
\author{P. Esquinazi}\email{esquin@physik.uni-leipzig.de}
\affiliation{Division of Superconductivity and Magnetism, Institut
f\"ur Experimentelle Physik II, Universit\"{a}t Leipzig,
Linn\'{e}stra{\ss}e 5, D-04103 Leipzig, Germany}

\begin{abstract}
Within the BCS theory of superconductivity we calculate the
superconducting gap at zero temperature for metallic
hydrogen-graphene system in order to estimate the superconducting
critical temperature of quasi two dimensional highly oriented
pyrolytic graphite. The obtained results are given as a function of
the hydrogen-induced density of carriers $n$ and their effective mass
$m^\star$. The obtained gap shows a Maxwell-like distribution with a
maximum of $\sim 60~$K at $n \sim 3 \times 10^{14}~$cm$^{-2}$ and
$m^\star/m = 1$. The theoretical results are discussed taking into
account recent experimental evidence for granular superconductivity
in graphite.
\end{abstract}
 \maketitle

\section{Introduction}
Since the discovery of high temperature superconductivity
\cite{ber86} we learned that low dimensional structures are good
candidates for high critical superconducting temperature $T_c$
\cite{ott08}. Related to this dimensionality effect it appears
appropriate to mention also the role played by two dimensional (2D)
interfaces in triggering superconductivity in nominally
non-superconducting environment \cite{mun06,mun08,goz08}. One of the
main paradigms for 2D carrier systems, graphene as well as highly
oriented pyrolytic graphite (HOPG) or multigraphene have attracted
considerable attention recently, partially due to the possibility of
regulating the carrier density $n$ by the field effect
\cite{novoscience,kim05}. On the other hand, hints for the existence
of superconductivity in HOPG have been published in the last years
\cite{kopejltp07,esq08}.

Let us first point out some interesting aspects of multigraphene that
should play a role in triggering the superconductivity phenomenon.
Whereas graphene should be nominally a 2D system, multigraphene or
HOPG is a quasi-2D system due to the weak coupling between graphene
layers, which is the reason for the huge anisotropy in resistivity,
for example \cite{yakovadv03}. Both materials, graphene as well as
HOPG have high energetic phonons going up to $\sim 0.15$~eV with many
other branches at lower energies \cite{nic73,sib97}. High energy
phonons due to small element mass provide an excellent condition for
superconductivity as in the case of metallic H at high pressures
\cite{ash68}. From this point of view it seems reasonable to look for
superconductivity in these materials taking into account the whole
phonon spectral function weighted by the electron-phonon coupling
\cite{ben08}.

For HOPG it has been recently shown that upon temperature (and defect
concentration) $n \sim 10^8 \ldots 10^{12}~$cm$^{-2}$.~\cite{gar08}
The lowest value obtained in clean samples and at low temperatures is
clearly smaller than in free-standing as well as fixed-on-substrate
graphene samples. To achieve a clear increase in the carrier density
the case of attached hydrogen in HOPG is of special interest.
According to theoretical results upon the amount and the way hydrogen
is attached to a graphene layer, a semiconducting, metallic or even a
magnetically orderer state appears \cite{yaz07,duplock04,pis07}. In
case of metallization the electron density may increase dramatically
in the region near hydrogen, e.g. 0.1 to 1 electron per unit cell.
Added to this effect, the effective mass increases from the usual
very small values $m^\star \lesssim 0.01 m $ to nearly the free
electron mass $m$ \cite{duplock04}. Note that in this case no Dirac
dispersion relation is valid for graphite carriers but the usual
quadratic one,  dispersion that we will assume through all this work.
The hydrogen-graphene bound system is interesting due to the
influence in the spectral density similarly to the case of
superconductivity in aromatic molecules \cite{ben08,kre90}.

On the other hand, we know that the reaction of hydrogen adsorbed on
carbon is endothermic and difficult to realize. However, experimental
results indicate large amounts of hydrogen is present in HOPG samples
\cite{rei06}. Hydrogen in HOPG might be included through the
synthesis of this material, obtained after high-pressure and
high-temperature treatment of polymers (e.g. kapton) foils. We expect
therefore hydrogen can be bounded to carbon, specially around
vacancies and other defects, or at the interfaces conforming regions
or pockets with different electron densities. Hydrogen bounded at
defects might be the origin for the ferromagnetic properties
\cite{pabloprb02,barzola2},  the anomalous transport observed in HOPG
in the last years \cite{yakovadv03} and for the existence of
non-homogeneous, granular-like superconductivity \cite{esq08},
possible at the interfaces between highly crystalline graphite
regions in HOPG samples \cite{bar08}.

The purpose of this work is to calculate the superconducting energy
gap at $T = 0$~K for a graphene layer and extend this result to the
anisotropic graphite case. Following the Mermin-Wagner theorem in a
pure 2D system there is no superconductivity \cite{mer66} and
therefore the problem should be treated as a
Beresinskii-Kosterlitz-Thouless (BKT) phase transition
\cite{ber70,kos73}. In the anisotropic case the problem is solved
having a small anisotropy $\epsilon$, which resembles the low
coupling between graphene layers in graphite. As in the 2D
anisotropic Heisenberg model, no matter how small is $\epsilon$, one
has an appreciable critical temperature because of the logarithmic
behavior of the thermal fluctuation influence \cite{lev92,bre76}. The
results of this work indicate that hydrogen could play a decisive
role triggering high temperature superconductivity in graphite.

\section{Superconductivity at high metallic density of
carriers}

The estimate of the critical temperature $T_c$ using BCS is done from
the energy gap equation given by \cite{de99,bog59}
\begin{equation}\label{Delta}
    \Delta(E) = -N(0) \int \textrm{d}E' V(E-E') \Delta(E') (1 -
    2f(E'))/2E' \,,
\end{equation}
where $N(0)$ is the density of states at the Fermi energy $E_F$,
$V(E-E') = V_P(E-E') + V_C$ is the interaction potential that we
split into the electron-phonon term and the Coulomb potential, this
last taken as a constant for a given carrier density. The potential
$V_P(E-E')$ depends on the pair-interaction-assisted phonon energy
$E-E'$. Both potentials will be estimated below. Finally the
Fermi-Dirac distribution function $f(E')$ at a given $T$ that we set
at $T_c$.

At high metallic densities $E_F \gg E_D$ (this last the Debye energy)
$T_c$ is estimated assuming that electron pairs are formed with an
energy difference up to $E_D$ around $E_F$. Then, the well known
equation for the zero temperature energy gap (weak coupling limit)
\cite{ben08,de99,bog59}
\begin{equation}\label{Delta2}
    \Delta(T = 0) \approx 2 E_D \exp(-1/(\lambda-\mu^\star))\,,
\end{equation}
is obtained with $\lambda = N(0)<V_P>$ and
\begin{equation}\label{mu1}
 \mu^\star = \mu/(1+\mu \ln(E_F/E_D))\,,
 \end{equation}
with $\mu = N(0)<V_C>$ where the $<...>$ means the average within
$E_F$ or $E_D$ and the screening length of the corresponding
potentials. The well known result (\ref{Delta2}) is obtained solving
Eq.~(\ref{Delta}) by introducing cutoffs energies for electrons and
phonons. Using an Einstein approximation Morel and Anderson obtained
the same results  for $\Delta(0), \lambda$ and $\mu^\star$ without
introducing cutoffs \cite{mor62}. The results remain the same for a
2D system but due to the BKT phase transition it is valid only at $T
= 0~$K.

\section{Estimate of $\Delta(0)$ in 3D and 2D as a function of carrier
density $n$}

We discuss now the differences between 2D and 3D. We express all the
necessary quantities as a function of $r_s$, the distance between the
carriers in units of the Bohr radius. The quantities $q_F(r_s) =
1.91/(a_B r_s)$ and $q_T(r_s) = 1.38/(a_B\sqrt{r_s})$ are the 3D
values for the Fermi vector and the inverse of the Thomas-Fermi decay
length. Analogously for 2D we have $\sqrt{2}/(r_s a_B)$ and $2/a_B$.
Notice that the inverse decay length in 2D does not depend on the
density $n = (\pi)^{-1} (a_B r_s)^{-2}$. Another important difference
is in the Coulomb interaction that behaves as
\begin{equation}\label{Vc3}
    V_C(q) = \frac{4\pi e^2}{q^2 + q_T^2(r_s)}\,,
\end{equation}
for 3D and
\begin{equation}\label{Vc2}
V_C(q) = \frac{e^2}{2(q + q_T(r_s))}\,,
\end{equation}
for 2D, where $q$ is the wave vector. These potentials should be
reliable for $r_s \lesssim 15$ above which Pines \cite{pin55} noted
 that convergence problems in the estimates may occur. Note that the value
$r_s = 15$ is still smaller than those needed for the Wigner crystal
formation \cite{wig34,tan89}.

In the 3D case the values of $\lambda$ and $\mu$ in
Eqs.~(\ref{Delta2}) and (\ref{mu1}) are given by
\begin{eqnarray}
  \lambda &=& \frac{r^2}{2(1+r^2)}\,, \\
  \mu &=& \frac{r^2\ln((1+r^2)/r^2)}{2}\,,
\end{eqnarray}
where $r^2 = (q_T(r_s)/2q_F(r_s))^2$. The Eqs.~(6,7) correspond
exactly to Eqs.~(46,44) from Ref.~\onlinecite{mor62}. Note that our
Eq.~(7) is the average angular value of Eq.~(26) of
Ref.~\onlinecite{mor62} that missed a factor $r^2$ in the
corresponding Eq.~(44). For the case of 2D we have $\mu$ as the
average angular of the Coulomb potential
\begin{eqnarray}
  \mu &=& \frac{q_T}{4q_F} \int_0^{2q_F} \frac{1}{q+q_T}dq \\
  &=& \frac{q_T}{4q_F} \ln \left(\frac{2q_F + q_T}{q_T}\right )\,,
\end{eqnarray}
and $\lambda$ is calculated as the average angular value of the
square Coulomb potential multiplied by $0.66$ because of  averaging
the $q-$moment. Therefore we have
\begin{eqnarray}
  \lambda &=& 0.66 \left ( \frac{q_T}{2q_F}\right )^2 \int_0^1 \left ( \frac{1}{(q_T/2q_F) + x} \right )^2 dx\\
  &=& 0.66 \frac{q_T^2}{2q_F} \left ( \frac{1}{q_T} - \frac{1}{q_T + q_F} \right )\,.
\end{eqnarray}
Now we are prepared to estimate the values of $\Delta(0)$ in 2D and
3D for different values of $r_s$, a function of the electronic
density. It should be notice that $\lambda$ and $\mu$ obtained as
defined previously agree with the values obtained by Morel and
Anderson \cite{mor62}. They obtained them including an Einstein
effective frequency as approximation for the phonon structure. Within
a Debye model similar equations are obtained but the phonon
contribution is taken as an average on the whole phonon spectra.

\begin{figure}[]
\begin{center}
\includegraphics[width=88mm]{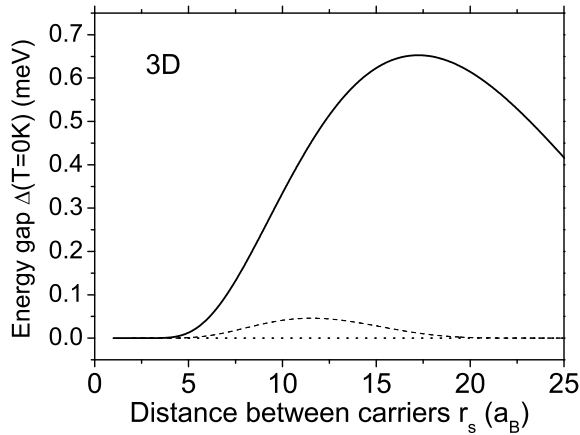}
\caption[]{Energy gap at $T = 0~$K vs. the distance between carriers
$r_s$ in units of Bohr radius $a_B = 4 \pi \epsilon_0 \hbar^2/m^\star
e^2$ for the 3D case and for the effective masses $m^\star = 3, 1,
0.1$ of the free electron mass (continuous, dashed and dotted lines,
respectively). The curves were obtained using
Eqs.~(\protect\ref{Delta2},3,4,6,7) and the parameter $E_D = k_B
860~$[K] as an average over all frequencies.} \label{Fig1}
\end{center}
\end{figure}

We found that $\Delta(0)$ is in general about $10 \ldots 100$ times
larger in 2D than in 3D at similar values of $r_s$. In Fig.~1 we
present $\Delta(0)$ vs. $r_s$ for the 3D case for 3 values of
$m^\star$ corresponding to 0.1, 1 and 3 the free electron mass.
Figure 2 shows the results for the 2D case. What can be the physical
reasons for having energy gaps much larger in 2D than in 3D? They are
related to the different weights and values of the ratios $q_T/2q_F$
and $q_T/q_D$ ($q_D$ is the averaged maximum phonon wave vector).
Their contributions in 2D are dominant and depend on
 the effective mass, as seen in Fig.~3 where we plot
the values of $q_T/2q_F$ and $q_T/q_D$, with $q_D = 10^8~$cm$^{-1}$
for 2D (as in 3D). Note that the differences in $\Delta(0)$ between
the 2D and 3D cases are larger the smaller the effective mass
$m^\star$.

\begin{figure}[]
\begin{center}
\includegraphics[width=88mm]{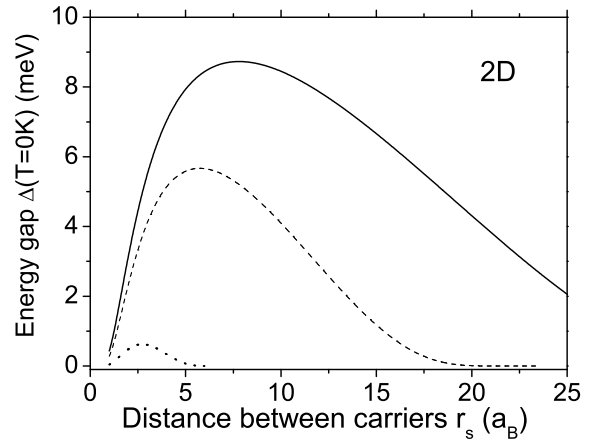}
\caption[]{The same as in Fig.~1 but for the 2D case with the same
parameters.} \label{Fig2}
\end{center}
\end{figure}

\begin{figure}[]
\begin{center}
\includegraphics[width=88mm]{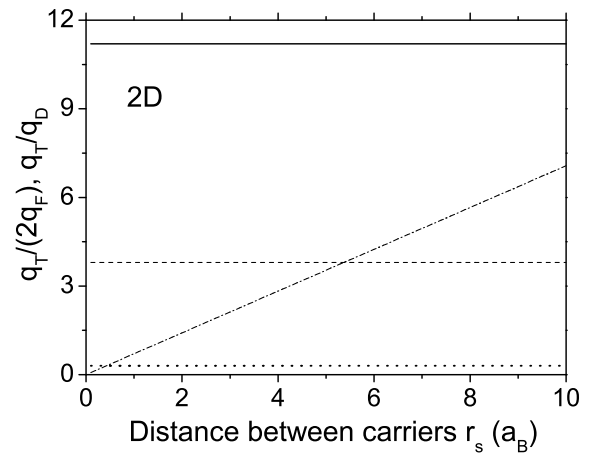}
\caption[]{Thomas-Fermi wavevector ratios $q_T/2q_F = r_s/\sqrt{2}$
(dashed-dotted line) and $q_T/q_D$ for the 2D case for the three
different effective masses 3, 1 and 0.1 of the free electron mass
(continuous, dashed and dotted lines).} \label{Fig3}
\end{center}
\end{figure}

It is adequate to study quasi 2D problems with a degree of anisotropy
provided by the ratio between the conductance in the atomic lattice
plane and perpendicular to it, i.e. $\epsilon \simeq
\sqrt{\sigma_{c}/\sigma_{ab}}$. For example, in the case of HOPG we
have an in-plane conductance $\gtrsim 10^4$ than the conductance
perpendicular to the plane. As in the case of the Heisenberg model in
2D with anisotropy, even if it is small, this is very effective to
recover again the superconducting phase transition at $T
> 0~$K by a logarithmic function of this anisotropy, i.e. $T_c
\propto 1/\ln(K(1-\epsilon)) \propto -1/\ln(\epsilon)$ for $\epsilon
\rightarrow 0$, where $K(x)$ is the elliptic function
\cite{lev92,bre76}. The anisotropy produces a variation of $T_c$ of
$\sim 10-20\%$ of the value obtained from the simple estimate using
the BCS result $T_c \sim 2\Delta(0)/(3.5 k_B)$. Therefore the value
of $\Delta(0)$ remains a good approximation for the anisotropic quasi
2D case.

We performed first calculations for HOPG at relatively low
densities. We note that in graphene as well as in HOPG the
expected $T_c$
 is negligible
small at electron densities $n < 10^{13}~$cm$^{-2}$. According to the
BCS approach and our estimates neither materials can be
superconductors at  those electron densities. However, for the
electron densities obtained for the case of hydrogen fixed in special
(e.g. Stone-Wales) defects we have $n \sim 10^{14} \ldots 3 \times
10^{14}~$cm$^{-2}$ with a metallic-like band character that crosses
the Fermi level with $m^\star \sim m$ \cite{duplock04}. The estimated
critical temperatures for these densities are around 25~K, values
comparable to the observations in multigraphene samples
\cite{esq08,bar08}. Note that decreasing $n$ by a factor of two
increases $T_c$ approximately by a similar factor, see Fig.~2.
According to our estimate, a further decrease of $n$, however, would
produce a decrease of  the critical temperature, see Fig.~2.

Although defects in the graphene/graphite structure would increase
the carrier density in a first stage, we believe that attached
hydrogen may be the most probable reason for the local increase of
$n$. The idea that hydrogen may trigger superconductivity in graphite
due to the large increase in the electron density provides also a way
to understand several experimental facts that we summarize below.
Experimental results from SQUID \cite{yakovjltp00,kope06,kopejltp07}
and magnetotransport \cite{yakovadv03,esq08} indicate that the
possible superconducting state in HOPG has granular character:
neither percolation in transport nor Meissner effect are observed so
far for samples larger than $\sim 10~\mu$m. Irreversibilities in the
magnetotransport of mesoscopic multigraphene samples \cite{esq08} are
compatible with the existence of superconducting patches connected by
semiconductor regions, similar to those irreversibilities observed in
granular high-temperature superconductors \cite{fel03}. Transport
measurements at different regions of the same micrometer small and a
few tens of nanometers thick multigraphene samples reveal an
inhomogeneous behavior compatible with the existence of metallic- and
semiconductor-like regions \cite{esq08}. Compatible with this view of
HOPG, electric field force microscopy (EFM) measurements on its
surface indicate the existence of large variations in the electronic
potential providing a clear hint that HOPG is an inhomogeneous
electronic system \cite{lu06}. A recently done correlation between
the thickness dependence of the resistivity and magnetotransport with
the internal microstructure of HOPG suggests that the internal
interfaces between crystalline graphite regions are the regions where
superconductivity can be located \cite{bar08}. These interfaces may
have enough hydrogen trapped to increase the electronic density,
triggering superconductivity in different regions with different
critical temperatures, keeping the quasi-two dimensionality in
agreement with the large anisotropy observed in the experiments
\cite{esq08}.

We should mention other approaches that deal with superconductivity
in graphene and graphite. Starting with graphene \cite{uch07,kop08}
we would like to note that for a carrier density $n <
10^{13}~$cm$^{-2}$ the critical temperature $T_c < 10^{-5}~$K.
Therefore, one needs to increase drastically the electron
concentration in order to have a $T_c$ in the few Kelvin range.
Another publication \cite{doni07} obtains very large numbers for
$T_c$ in graphite. The basis of this result is to add certain atoms
to graphite, in particular sulfur. The attractive interactions
between electrons are based in electronic correlations effects. It is
basically the same type of approach that we use but with phonons. It
is clear that to produce superconductivity in graphite one has to
incorporate a material into its planes to reise the carrier density
to $\sim 10^{14}~$cm$^{-2}$. Otherwise we find that no appreciable
superconductivity can be expected in graphite within the mean field
approach.

\section{Conclusion}
In conclusion, the superconducting-like behavior observed in bulk
HOPG as well as in mesoscopic multigraphene samples could be
explained applying the known techniques within the BCS approach and
calculating $\Delta(0)$ for the quasi-2D problem. From our results we
expect clear variations of $T_c$ with the electronic density,
indicating that neither perfect HOPG nor perfect graphene could be
superconducting. We have presented results that indicate a huge
increase in $T_c$ for the 2D case with respect to the 3D for the same
set of parameters. This is due to the behavior of $q_T(r_s)$ and
$q_F(r_s)$ providing much higher values for $\Delta(0)$ in the 2D
case. We have applied the BCS approach to the system
graphene-hydrogen and have found that $\Delta(0) \gtrsim 25~$K are
possible with reasonable electronic densities. Upon effective mass
and electronic density, high critical temperatures in HOPG-hydrogen
system may be realized and therefore the problem of superconductivity
in graphite should be taken with more attention.

We gratefully acknowledge the  support of the DAAD under Grant No.
D/07/13369 (``Acciones Integradas Hispano-Alemanas"). One of us
(P.E.) acknowledges discussions with Y. Dagan.  One of us (N.G.) is
supported by the Leibniz Professor fellowship of the University of
Leipzig.


\end{document}